\documentclass[preprint,aps,pre,superscriptaddress,citeautoscript,floatfix,longbibliography]{revtex4-1}
\usepackage{graphics,graphicx,mathrsfs,color,amsmath,appendix,framed,tcolorbox,bm,enumerate}
\usepackage{hyperref}
\usepackage{empheq}
\usepackage{cases}
\newcommand{\ud}{\mathrm{d}}

\newcommand{\change}[1]{\textcolor{black}{#1}}

\begin{document}

\title{Wetting equilibrium in a rectangular channel}
\author{Tian Yu}
\affiliation{State Key Laboratory of Hydroscience and Engineering, Tsinghua University,
Beijing 100084, China}
\author{Qicheng Sun}
\affiliation{State Key Laboratory of Hydroscience and Engineering, Tsinghua University,
Beijing 100084, China}
\author{Chen Zhao}
\affiliation{Center of Soft Matter Physics and its Applications, Beihang University, Beijing 100191, China}
\affiliation{School of Physics, Beihang University, Beijing 100191, China}
\author{Jiajia Zhou}
\email[]{jjzhou@buaa.edu.cn}
\affiliation{Center of Soft Matter Physics and its Applications, Beihang University, Beijing 100191, China}
\affiliation{School of Chemistry, Key Laboratory of Bio-Inspired Smart Interfacial Science and Technology of Ministry of Education, Beihang University, Beijing 100191, China}
\author{Masao Doi}
\affiliation{Center of Soft Matter Physics and its Applications, Beihang University, Beijing 100191, China}

\begin{abstract} 
When a capillary channel with corners is wetted by a fluid, there are regions where the fluid fills the whole cross-section and regions where only the corners are filled by the fluid. 
The fluid fraction of the partially-filled region, $s^*$, is an important quantity related to the capillary pressure. 
We calculate the value of $s^*$ for channels with a cross-section slightly deviated from a rectangle: the height is larger in the center than those on the two short sides. 
We find that a small change in the cross-section geometry leads to a huge change of $s^*$. 
This result is consistent with experimental observations.
\end{abstract}

\maketitle

\section{Introduction}

Wetting and drying in porous media are ubiquitous phenomena in our daily life \cite{Bear, deGennes1985, Bonn2009}.
They are also important in many applications, including environmental science, oil recovery, and processes in food, textile, and pharmaceutical researches. 
Porous media have complicated structures involving wide distribution of pore size and various junction geometries, which make the comprehensive understanding difficult. 
Simplified structures have been used to understand the fluid transportation in porous media, such as capillary tubes with corners \cite{Prat2007}.
One beautiful example is the study of the evaporation in a capillary tube \cite{Chauvet2009}, where the drying dynamics in a simple square tube resembles that in a porous medium. 

An important quantity governing the capillary phenomena in porous media is the capillary pressure $p_c$ \cite{Bear}, defined as the pressure difference across the interfaces separating two phases (normally one is air and the other one is the fluid). 
If the shape of the meniscus is known, the capillary pressure is given by the Laplace pressure
\begin{equation}
  p_c = \gamma \left( \frac{1}{r_1} + \frac{1}{r_2} \right) = \frac{ 2 \gamma}{r^*},
\end{equation}
where $\gamma$ is the surface tension of the fluid, $r_1$ and $r_2$ are the meniscus' two principal radii of curvature , and $r^*$ is the mean radius of curvature ($2/r^* = 1/r_1 + 1/r_2$).
In a capillary tube with corners, the fluid can either fill up the whole cross-section or only part of the cross-section, with a transition region connecting them (see Fig.~\ref{fig:sketch}). 
At equilibrium, the capillary pressure is equal at every point of the meniscus, one then has the freedom to choose where to calculate the capillary pressure. 
It turns out that the calculation is easier at the partially filled region because one of the principal curvature (along the tube axis) is zero. 
This is the method proposed first by Mayer and Stowe \cite{Mayer1965} and Princen \cite{Princen1969, Princen1969a, Princen1970}, and extended later by Mason and Morrow \cite{Mason1984}. 
One related quantity is the saturation of the partially filled region, which we will describe as $s^*$. 
The relation between $p_c$ and $s^*$ will be given later [Eq.~(\ref{eq:pc})].
In this work, we shall discuss the value of $s^*$ for a capillary channel with a cross-section slightly deviated from a rectangle. 

\begin{figure}[ht]
  \includegraphics[width=0.9\columnwidth]{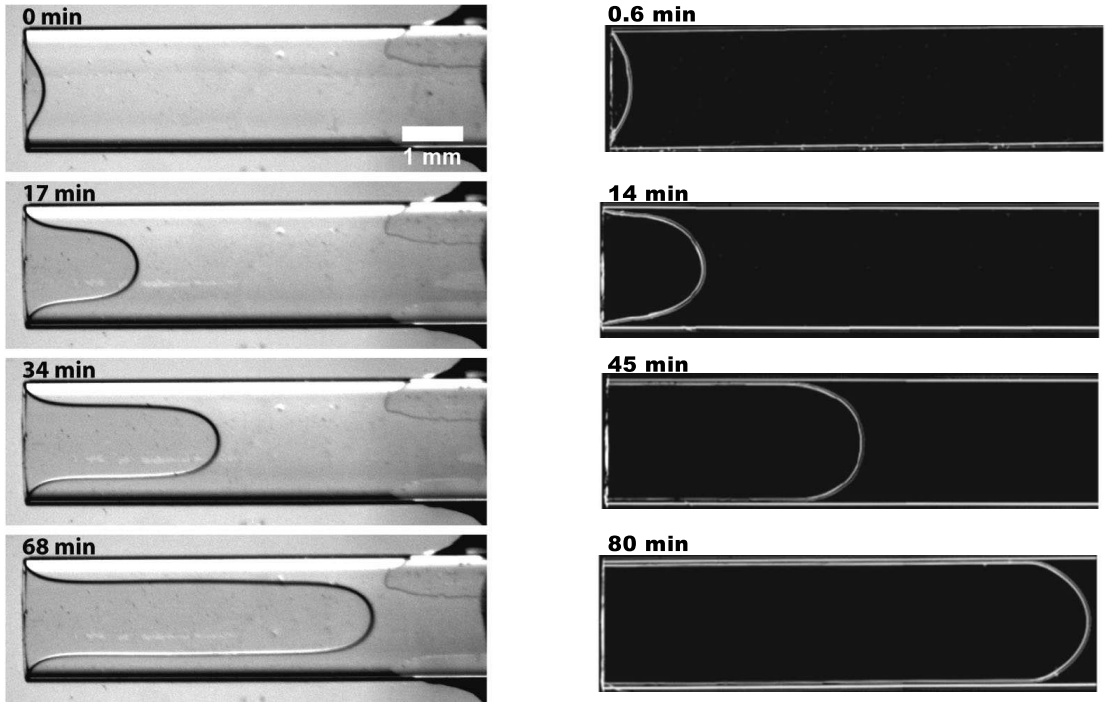}
  \caption{Top views of the nearly-rectangle channel (open to the left) at different drying time. The left column is from Ref.~\cite{Keita2016}, in which the height of the cross-section is large in the middle. The right column is from Ref.~\cite{Seck2018}, in which the cross-section is a nearly-perfect rectangle. (Reprinted from Refs.~\cite{Keita2016} and \cite{Seck2018} with permission from the Springer Nature.)}
  \label{fig:exp}
\end{figure}

Two recent studies in rectangular channels motivated our study. 
Keita \emph{et al.} \cite{Keita2016} studied the drying dynamics in a nearly-rectangular channel, the thickness of which is slightly larger in the middle than in the edges (the thickness is 115 $\mu$m in the middle and 95 $\mu$m at the edges; the width of the channel is 2 mm). 
During the evaporation, two thick fluid fingers were present in the partial-saturated region (left column in Fig.~\ref{fig:exp}). 
This is in contrast to a later study by Seck \emph{et al.} \cite{Seck2018}. 
In this case, the rectangular channel was nearly perfect, and very thin fluid columns were observed (right column in Fig.~\ref{fig:exp}). 
These two experiments clearly showed that slight deviation from the rectangular geometry changes the value of $s^*$ enormously.
The purpose of this paper is to explain these phenomena based on a free energy model. 

\section{Model}

\begin{figure}[ht]
  \includegraphics[width=0.6\columnwidth]{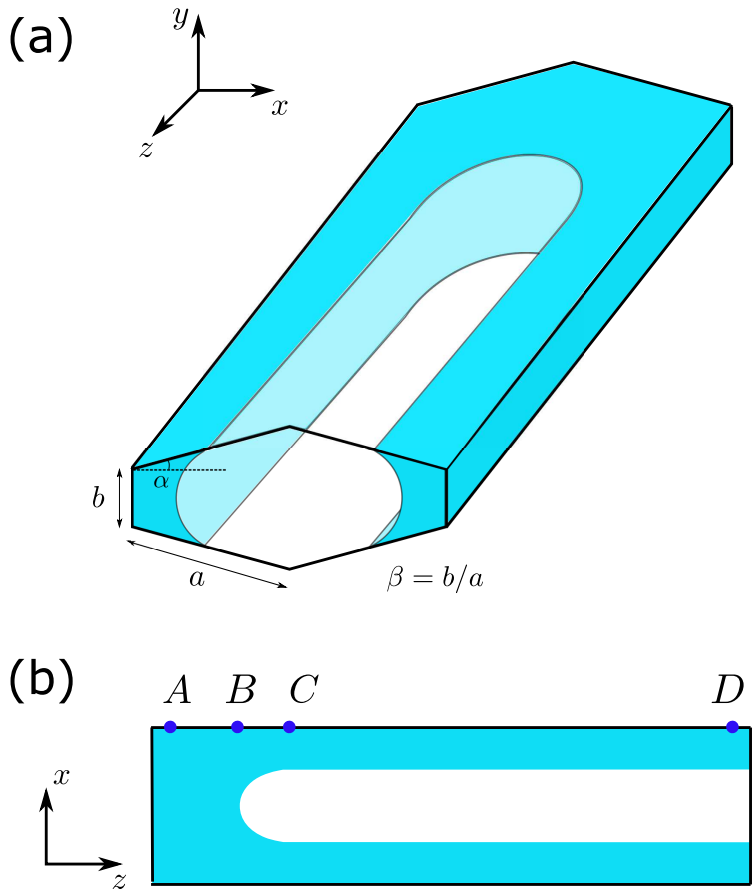}
  \caption{(a) Schematic picture of partially-filled channel with a cross-section slightly deviated from a rectangle. The cross-section is characterized by the top/bottom edge $2a$, side edge $b$, leading to an aspect ratio $\beta=b/a$. 
These two sides forming an angle that is slightly larger than the normal angle by $\alpha$. 
The channel is filled partially by a fully wetting fluid with the equilibrium contact angle $\theta_E=0$. 
(b) The view from the cutting $x$-$z$ plane. $AB$ part is fully filled by the fluid and $CD$ is partially filled, with a transition region $BC$ in between.
In these pictures, the channel is sealed by a wetting material at the end near $A$ and a neutral material at the end near $D$.}
  \label{fig:sketch}
\end{figure}

We consider a capillary channel with a cross-section slightly deviated from a rectangle.
The channel is horizontally placed along the $z$-direction, as illustrated in Fig.~\ref{fig:sketch}(a). 
The total length of the long side is $2a$ and the short side of the cross-section has a length of $b$.
For convenience, we define a length ratio $\beta = b/a$, and we shall focus on the situation when this ratio is small, $\beta<1$.  
The two sides ($a$ and $b$) form an angle of ${\pi}/{2} + \alpha$, which is slightly larger than the normal angle. 
We focus on the situation that $\alpha$ is small, corresponding to the case that the cross-section is quite close to a rectangle.

Now we place some wetting fluid into the channel and seal both ends. 
The amount of fluid is chosen such that the fluid only partially fills the channel, as shown schematically in Fig.~\ref{fig:sketch}(a).
In the thermodynamic equilibrium, some portion of the channel is fully filled by the fluid with the saturation $s=1$ (the saturation $s$ is defined as the area fraction occupied by the fluid in the $x$-$y$ cross-section), and some portion is only partially filled with a saturation $s^*<1$. 
Our goal is to determine the value of $s^*$, and how its value varies when other geometrical parameters, such as $\alpha$ and $\beta$, change.

The coexistence between the fully-filled region and partially-filled region can be seen more clearly in the cutting $x$-$z$ plane shown in Fig.~\ref{fig:sketch}(b): $AB$ part corresponds to fully saturation $s=1$, while $CD$ part corresponds to the partial saturation $s^*$. 
We shall refer the fully-saturated region ``bulk'' and the partially-saturated region ``finger''.
There is a transition region $BC$ going from the bulk to the finger.
The length of the transition region is finite and with a characteristic size of $a$. 
The effect of the transition zone can be neglected in the thermodynamic limit when the length of the bulk ($AB$) and the finger ($CD$) both become larger than the length of the transition zone ($BC$) \cite{Weislogel2011, Weislogel2012}.
The fluid has a surface tension $\gamma$ and an equilibrium contact angle $\theta_E$. 
In the cornered geometry, only when the condition 
\begin{equation}
  \label{eq:ConcusFinn}
  \frac{1}{2} \left( \frac{\pi}{2} + \alpha \right) + \theta_E  < \frac{\pi}{2}
\end{equation}
is fulfilled, the partially-filled portion is possible \cite{Concus1969}. 
In this study we focus on the fully-wetted fluid, i.e., the contact angle $\theta_E=0$. 
For the fully-wetted fluid, the above condition (\ref{eq:ConcusFinn}) is satisfied when $\alpha < \change{\pi/2}$. 
Since we are interested in the case $\alpha$ is small, the condition (\ref{eq:ConcusFinn}) is automatically satisfied. 
We assume the channel height $b$ is less than the capillary length, so the effect of the gravity can be neglected.
The smallest dimension of the channel (also $b$ here) should be larger than a few nanometers, thus the fluid surface is well-defined.
The only contribution to the energy is the interfacial energy.

\section{Free energy density}  

\begin{figure}[ht]
  \includegraphics[width=0.8\columnwidth]{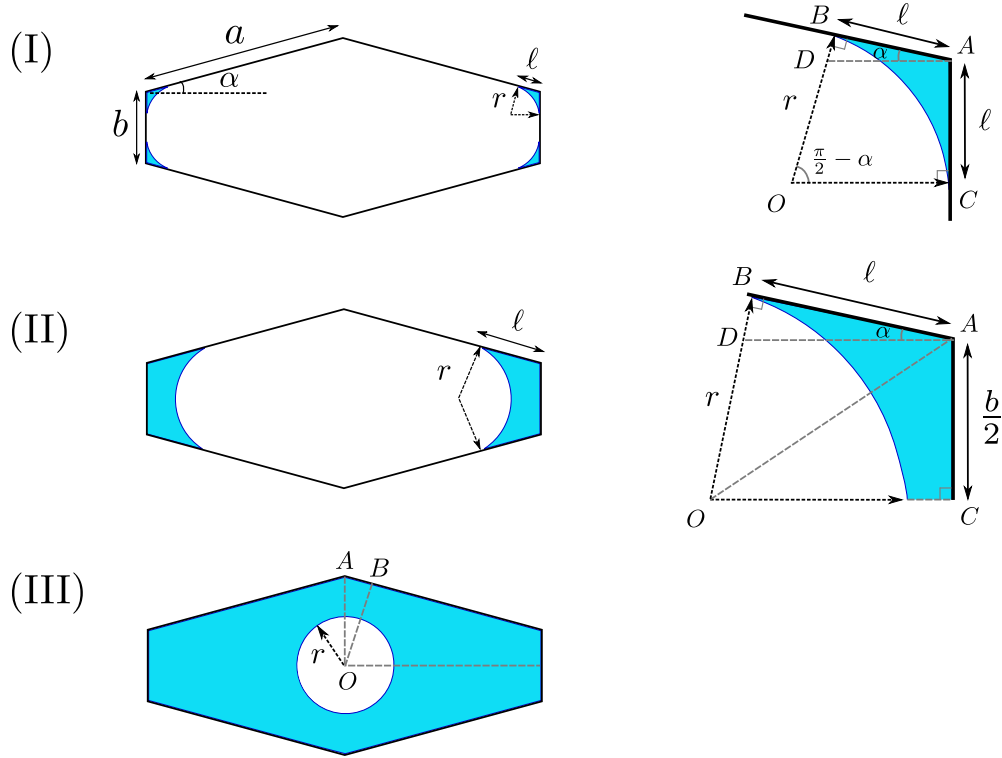}
  \caption{Three scenarios of the wetting liquid in the cross-section of a nearly-rectangular channel. 
(I) The liquid-vapor interfaces form four separated arcs in the corners. 
(II) The liquid-vapor interfaces form two separated arcs in the corners, with $b$ side is fully covered by the fluid.
(III) The liquid-vapor interface is a full circle, with the perimeter of the cross-section are all covered by the fluid.}
  \label{fig:rect}
\end{figure}

In the $x$-$y$ plane, the liquid-vapor interface of the meniscus in the finger region is part of a circle. 
Possible shapes of the meniscus are illustrated in Fig.~\ref{fig:rect}. 
We define $f(s)$ as the free energy density of the wetting liquid as the free energy per
unit length along the $z$ axis, which can be expressed as a function of the saturation $s$. 
For the nearly-rectangular channel, we distinguish three cases of wetting scenarios, based on whether the vapor-solid interfaces exist and how the contact lines intersect the side walls.

\subsection{Case (I)}

The situation is shown in Fig.~\ref{fig:rect}(I). 
The liquid-vapor interfaces are composed of four separate arcs in the corners. 
There are solid surfaces exposed to the vapor in both $a$ side and $b$ side. 
Since the value of $\alpha$ is small, we assume that the liquid doesn't wet in the two corners constructed by the two adjacent $a$ sides.   
The radius of curvature of the liquid-vapor interface is denoted by $r$, and $\ell$ is the wetting length on the $a$ side. 
The radius $r$ can be expressed as a function of $\ell$
\begin{equation}
  \label{eq:r_case1}
  r = \overline{BD} + \overline{DO} 
    = \overline{AB} \tan\alpha + \overline{AC} / \cos\alpha 
    = \ell \left( \tan\alpha + \sec\alpha \right) .
\end{equation}
The area occupied by the fluid is the area of quadrilateral $ABOC$ minus the sector $OBC$ (counting 4 corners)
\begin{equation}
  S = 4 \left( \ell r - (\frac{\pi}{2} - \alpha) \frac{1}{2} r^2 \right)
    = 4 \ell^2 (\tan\alpha + {\sec\alpha} ) \left[ 1 - (\frac{\pi}{4} - \frac{\alpha}{2}) 
          (\tan\alpha + {\sec\alpha} ) \right] . 
\end{equation}
The total area is a constant
\begin{equation}
  S_0 = 2 \, a\cos\alpha \, b + 2 \, a\cos\alpha \, a \sin\alpha 
      = 2a^2 \cos\alpha ( \sin\alpha + \beta ),
\end{equation}
from which we can calculate the saturation $s$ 
\begin{equation}
  \label{eq:s_case1}
  s = \frac{S}{S_0} = \frac{ 2 (\tan\alpha + {\sec\alpha} )}{\cos\alpha (\sin\alpha + \beta) } 
      \left[ 1 - (\frac{\pi}{4} - \frac{\alpha}{2}) (\tan\alpha + {\sec\alpha} ) \right]
      \Big( \frac{\ell}{a} \Big)^2 .
\end{equation}

Since there are solid surfaces exposed to the vapor, the range of $\ell$ is limited by $\ell < b/2$. 
This leads to the following constraint on the saturation
\begin{equation}
  0 < s < s_{c1},
\end{equation}
\begin{equation}
  \label{eq:sc1}
  s_{c1} = s(\ell=\frac{b}{2}) 
    =  \frac{ (\tan\alpha + {\sec\alpha} )}{ 2 \cos\alpha (\sin\alpha + \beta) } 
    \left[ 1 - (\frac{\pi}{4} - \frac{\alpha}{2}) (\tan\alpha + {\sec\alpha} ) \right] \beta^2 .
\end{equation}

The free energy density $f(s)$ is expressed as
\begin{equation}
  \label{eq:fs-case1}
  f(s) = 8 \ell (\gamma_{\rm SL} - \gamma_{\rm SV}) + 4 (\frac{\pi}{2}-\alpha)r \gamma 
  = -8 \ell \gamma  + 4 (\frac{\pi}{2}-\alpha)r \gamma, 
\end{equation}
where $\gamma_{\rm SL}$ and $\gamma_{\rm SV}$ are the interfacial tensions between solid/liquid and solid/vapor, respectively. 
Here we have used Young's equation $\gamma \cos\theta_E = \gamma_{\rm SV} - \gamma_{\rm SL}$ and zero equilibrium contact angle $\theta_E=0$.
\change{We choose a dried surface as the reference point for the free energy. For the case of total wetting, there might be a precursor film of molecular thickness \cite{Tanner1979, deGennes1985, Bonn2009}. 
The first term in Eq. (\ref{eq:fs-case1}) is modified as $8 \ell (\gamma_{\rm SL} - (\gamma + \gamma_{\rm SL}))$, and the final expression remains valid.} 

Using Eqs.~(\ref{eq:r_case1}) and (\ref{eq:s_case1}), the free energy density can be written as a function of $s$
\begin{eqnarray}
  \frac{f(s)}{a\gamma} &=& -8 \left[ 1 - (\frac{\pi}{4}-\frac{\alpha}{2}) 
    (\tan\alpha + {\sec\alpha}) \right] \Big( \frac{\ell}{a} \Big) \\
  &=& - \sqrt{ \frac{ 32 \cos\alpha ( \sin\alpha + \beta)}{\tan\alpha + \sec\alpha}
      \left[ 1 - (\frac{\pi}{4} - \frac{\alpha}{2}) (\tan\alpha + {\sec\alpha} ) \right] } \sqrt{s}.
  \label{eq:f_case1}
\end{eqnarray}

\subsection{ Case (II)}

This situation is shown in Fig.~\ref{fig:rect}(II). 
Comparing to case (I), the two fingers on the $b$ sides merge into one finger.   
In this case, the vapor-solid interfaces only exist on the $a$ sides.
The radius $r$ is expressed as a function of $\ell$ as 
\begin{equation}
  \label{eq:r_case2}
  r = \overline{BD} + \overline{DO} 
    = \overline{AB} \tan\alpha + \overline{AC}/\cos\alpha 
    = \ell \tan \alpha + \frac{b}{2} \sec\alpha .
\end{equation}
The area occupied by the fluid is the summation of the triangle $ABD$ and trapezoid $ACOD$ minus the sector OB (counting 4 corners)
\begin{equation}
  S = 4 \left[ \frac{1}{2} \ell^2 \tan \alpha 
               + \frac{1}{2} \Big( {2\ell}\,{\sec\alpha} + \frac{b}{2} \tan\alpha \Big) \frac{b}{2} 
               - (\frac{\pi}{2} - \alpha) \frac{1}{2} r^2 \right].
\end{equation}
Using Eq.~(\ref{eq:r_case2}), the saturation $s$ is given by in term of $\ell$,
\begin{eqnarray}
  s = \frac{S}{S_0} &=& \frac{ \tan\alpha \left[ 1 - (\frac{\pi}{2} - \alpha) \tan\alpha \right]}
                   { \cos\alpha (\sin\alpha + \beta) } \Big( \frac{\ell}{a} \Big)^2 \nonumber \\
   && + \frac{ {\sec\alpha} \left[ 1 - (\frac{\pi}{2} - \alpha) \tan\alpha \right]}
                   { \cos\alpha (\sin\alpha + \beta) } \beta \Big( \frac{\ell}{a} \Big) 
      + \frac{ \tan\alpha - (\frac{\pi}{2} - \alpha) {\sec^2\alpha} }
              { 4 \cos\alpha ( \sin\alpha + \beta)} \beta^2
      \label{eq:s_case2}
\end{eqnarray}

The saturation $s$ in this case has two limits. 
The lower bound $s_{c1}$ is given by Eq.~(\ref{eq:sc1}) and the upper bound is constrained by $\ell < a$,
\begin{equation}
  s_{c1} \le s \le s_{c2}, 
\end{equation}
\begin{equation}
  \label{eq:sc2}
  s_{c2} = s(\ell=a) =  \frac{ (\tan\alpha + {\beta}\, {\sec\alpha})
    \left[ 1 - (\frac{\pi}{2} - \alpha) \tan\alpha \right] 
    + \frac{1}{4} \beta^2 \left[ \tan\alpha - (\frac{\pi}{2} - \alpha) {\sec^2\alpha} \right] }
    { \cos\alpha ( \sin\alpha + \beta)} .
\end{equation}

The free energy density $f(s)$ is expressed as
\begin{equation}
  f(s) = - 4 ( \ell + \frac{b}{2} ) \gamma + 4 (\frac{\pi}{2} - \alpha) r \gamma ,
\end{equation}
\begin{equation}
  \label{eq:f_case2}
  \frac{f(s)}{a\gamma} = - 4 \left[ 1 - (\frac{\pi}{2} - \alpha) \tan\alpha \right] \frac{\ell}{a} 
                         - 2 \left[ 1 - (\frac{\pi}{2} - \alpha) {\sec\alpha} \right] \beta .
\end{equation}
Here an explicit expression of the free energy density as a function of saturation is complicated, and it is easier to use $\ell$ as an intermediate variable. 
The range of $\ell$ is $ b/2 \le \ell < a$, leading to the two bounds $s_{c1}$ and $s_{c2}$ for the saturation.
The saturation and the free energy density as functions of $\ell$ are given by Eqs.~(\ref{eq:s_case2}) and (\ref{eq:f_case2}), respectively.

\subsection{Case (III)} 

This situation is shown in Fig.~\ref{fig:rect}(III). 
When the solid-vapor interfaces disappear, the liquid-vapor interface becomes a full circle of radius $r$.
The saturation is given as a function of $r$
\begin{equation}
  \label{eq:s_case3}
  s = 1 - \frac{\pi r^2}{S_0} = 1 - \frac{\pi}{2 \cos\alpha (\sin\alpha + \beta)} 
      \Big( \frac{r}{a} \Big)^2
\end{equation}
The radius is constrained by the shortest distance from the center to the $a$ side
\begin{equation}
  r \le \overline{OB} = \overline{OA} \cos\alpha =  ( a \sin\alpha + \frac{b}{2} ) \cos\alpha .
\end{equation}
This leads to a lower bound for the saturation
\begin{equation}
  s_{c3} < s \le 1, 
\end{equation}
\begin{equation}
  \label{eq:sc3}
  s_{c3} = 1 - \frac{\pi}{2} \cos\alpha 
      \frac{ (\sin\alpha + \frac{1}{2} \beta)^2 }{(\sin\alpha + \beta)} .
\end{equation}

The free energy density $f(s)$ is expressed as
\begin{equation}
  f(s)= -(4a+2b) \gamma + 2\pi r \gamma ,
\end{equation}
\begin{equation}
  \label{eq:f_case3}
  \frac{f(s)}{a\gamma} = - (4 + 2\beta) + 2 \sqrt{2\pi \cos\alpha ( \sin\alpha + \beta)(1-s)}.
\end{equation}

The free energy density at full saturation ($s=1$) is
\begin{equation}
  \label{eq:f_s1}
  \frac{f(1)}{a\gamma} = - (4 + 2 \beta) .
\end{equation}

\begin{figure}[ht]
  \includegraphics[width=1.0\columnwidth]{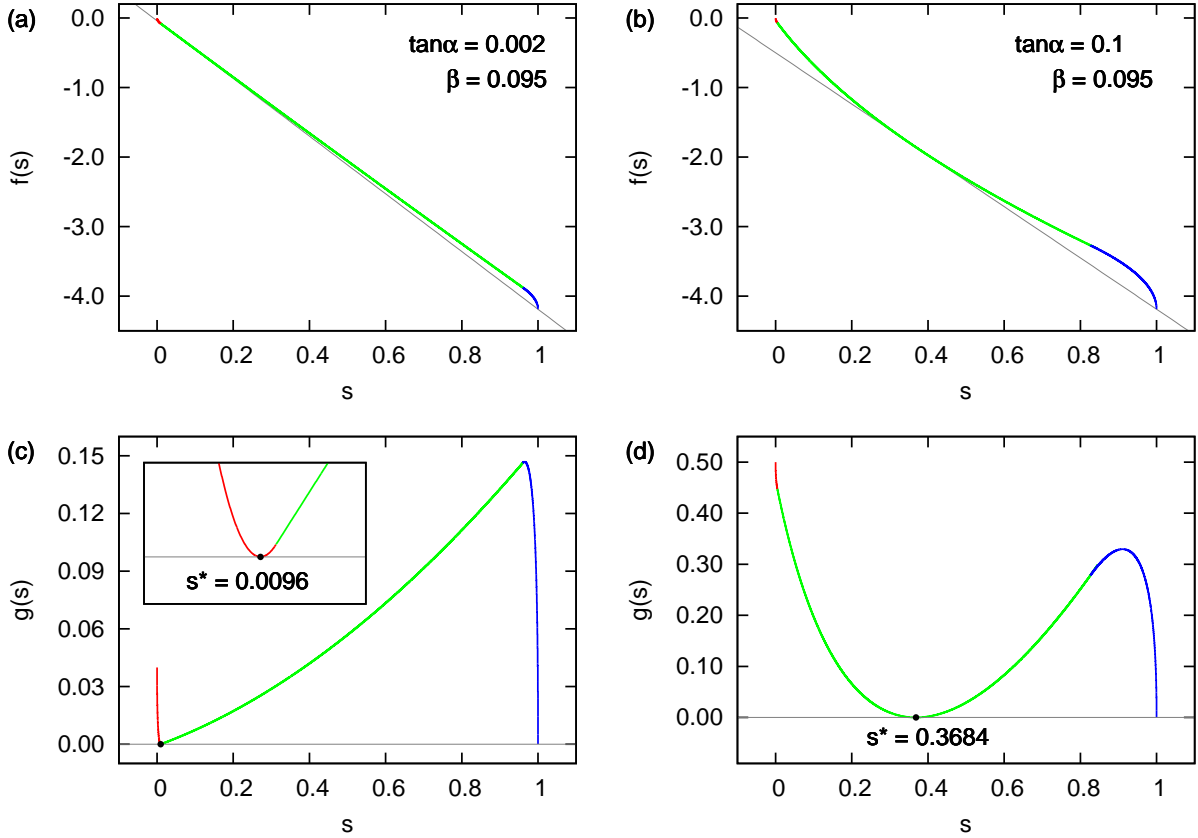}
  \caption{Example free energy density $f(s)$ as a function of the saturation $s$ for $\beta=0.095$. (a) $\tan\alpha = 0.002$. (b) $\tan\alpha = 0.1$. The different colors of the curves correspond to the three cases [red: (I), green (II), blue (III)]. The coexistence condition of (\ref{eq:tangential}) are shown as the gray lines. 
(c) and (d): the function $g(s)$ [Eq.~(\ref{eq:gs})] corresponding to (a) and (b). 
In (c), $s^*$ locates in the case (I) and in (d), $s^*$ locates in the case (II).
}
\label{fig:freeE}
\end{figure}

As a summary, the free energy density for the three cases are listed below  
\begin{numcases}{ \frac{f(s)}{a\gamma} = }
  - \sqrt{ \frac{ 32 \cos\alpha ( \sin\alpha + \beta)}{\tan\alpha + \sec\alpha}
    \left[ 1 - (\frac{\pi}{4} - \frac{\alpha}{2}) (\tan\alpha + {\sec\alpha} ) \right] } \sqrt{s} ,
  \quad 0 < s < s_{c1} \nonumber \\
  - 4 \left[ 1 - (\frac{\pi}{2} - \alpha) \tan\alpha \right] \frac{\ell}{a} 
           - 2 \left[ 1 - (\frac{\pi}{2} - \alpha) {\sec\alpha} \right] \beta ,
  \quad \quad s_{c1} \le s \le s_{c2} \nonumber \\
  - (4 + 2\beta) + 2 \sqrt{2\pi \cos\alpha ( \sin\alpha + \beta) } \sqrt{1-s} , 
  \quad \quad \quad \quad \quad \quad s_{c3} < s \le 1 \nonumber
\end{numcases}
Examples of free energy curves for $\beta=0.095$ are shown in Fig.~\ref{fig:freeE}(a) and (b) for $\tan\alpha = 0.002$ and $0.1$, respectively.
The aspect ratio $\beta$ takes the value in experiments of Ref.~\cite{Keita2016}.

\section{Saturation $s^*$} 

Once we have an expression of the free energy density as a function of saturation, we need a criterion to determine the partial saturation $s^*$. 
This will be derived in the following.

\subsection{Coexistence condition}

Suppose the total length of the channel is $L$ and the area of the cross-section is $S_0$. 
The total volume of the fluid is $V_l < L S_0 $. 
Along the channel, a fraction $\phi$ is partially saturated with a saturation $s<1$, and the remaining fraction $(1-\phi)$ is fully saturated by the fluid ($s=1$). 
Here we have neglected the contribution from the transition region.
The total free energy is given by 
\begin{equation}
  F_{\rm total} = L \phi f(s) + L (1-\phi) f(1). 
\end{equation}
To determine the thermodynamic equilibrium, we need to minimize the total free energy under the constrain that the total volume of the fluid is conserved
\begin{equation}
  \label{eq:Vl_conservation}
  L \phi S_0 s + L (1-\phi) S_0 = V_l.
\end{equation}

Introducing a Lagrangian multiplier $p_c$, we can minimize the following function $\mathscr{L}$ with respect to the fraction $\phi$ and the saturation $s$
\begin{eqnarray}
  \mathscr{L} &=& F_{\rm total} + p_c ( L \phi S_0 s + L (1-\phi) S_0 - V_l ) \nonumber \\
    &=& L \phi f(s) + L (1-\phi) f(1) + p_c ( L \phi S_0 s + L (1-\phi) S_0 -V_l )
\end{eqnarray}
The minimization leads to the following equations
\begin{eqnarray}
  \label{eq:dLdphi}
  \frac{\partial \mathscr{L}}{\partial \phi} &=& L \Big[ f(s) - f(1) + p_c S_0 ( s-1 ) \Big] = 0 ,\\
  \label{eq:dLds}
  \frac{\partial \mathscr{L}}{\partial s} &=& L \Big[ \phi f'(s) + p_c \phi S_0 \Big] = 0 ,
\end{eqnarray}
where the prime denotes the derivative with respect to the saturation, $f'(s) \equiv \ud f/\ud s$. 

The second equation (\ref{eq:dLds}) leads to the identification of the Lagrangian multiplier as the capillary pressure
\begin{equation}
  \label{eq:pc}
  p_c = - \frac{ f'(s) }{S_0} = - \frac{1}{S_0} \frac{ \ud f(s)}{\ud s} .
\end{equation}
Substituting the above expression into Eq.~(\ref{eq:dLdphi}), we obtain the condition for $s^*$
\begin{equation}
  \label{eq:tangential}
  \frac{ \ud f(s) }{ \ud s} \Big|_{s=s^*} = \frac{f(1)-f(s^*)}{1-s^*}.
\end{equation}
In Appendix \ref{sec:Mason}, we demonstrated that this condition is equivalent to the geometric condition proposed by Mason and Morrow in Ref.~\cite{Mason1984}.
Once the value of $s^*$ is determined, the partition of the fluid between the bulk and the finger is given by Eq.~(\ref{eq:Vl_conservation}).

Graphically, the coexistence condition (\ref{eq:tangential}) corresponds to drawing a straight line passing through the free energy curve at both points $(s^*, f(s^*))$ and $(1, f(1))$, and the line should also be tangential to the free energy curve at $s=s^*$. 
Example lines are shown as the gray lines in Fig.~\ref{fig:freeE}(a) and (b). 

Since the value of $f(s)$ changes a lot from $s=0$ to $1$, it is difficult to see the tangential line in the $f(s)$ plots. 
For the purpose of illustration, we define another function $g(s)$
\begin{equation}
  \label{eq:gs}
  g(s) = f(s) + (1-s) f'(s^*) - f(1) .
\end{equation}
This function has the properties that $g(s^*)=0$ and $g(1)=0$.
The condition (\ref{eq:tangential}) in the $g(s)$ plot then corresponds to a horizontal line passing through the origin. 
They are shown in Fig.~\ref{fig:freeE}(c) and (d). 

Since the free energy curves can be separated into three regions, the position of $s^*$ might be in each of those regions. 
However, for case (III), the free energy density (\ref{eq:f_case3}) has the form $f(s) = F_3 \sqrt{1-s} + {\rm constant}$ with $F_3 > 0$. 
The second derivative $f''(s) < 0$, so $s^*$ cannot exist in the range of case (III). 
We will discuss two cases that either $s^*$ is located in case (I) or case (II).

\subsection{Case (I)}

In this case, the free energy density (\ref{eq:f_case1}) has a form 
\begin{equation}
  \frac{f(s)}{{a\gamma}} = - F_1 \sqrt{s}, \quad 
  F_1 = \sqrt{ \frac{ 32 \cos\alpha (\sin\alpha + \beta)}{ \tan\alpha + \sec\alpha} 
    \left[ 1 - (\frac{\pi}{4}-\frac{\alpha}{2})(\tan\alpha + \sec\alpha) \right] }.
\end{equation}
Substituting the above expression into Eq.~(\ref{eq:tangential}) and using $f(1)$ from Eq.~(\ref{eq:f_s1}), we obtain an equation for $s^*$
\begin{equation}
  F_1 s^* - (8+4\beta) \sqrt{s^*} + F_1 = 0.
\end{equation}
The solution is 
\begin{eqnarray}
  s^* &=& \frac{1}{F_1^2} \Big( 4 + 2\beta - \sqrt{(4+2\beta)^2 - F_1^2} \Big)^2 \nonumber \\
  &=& \frac{ (\tan\alpha + \sec\alpha)  \left( (2+\beta) - \sqrt{ (2+\beta)^2 - \frac{8\cos\alpha (\sin\alpha + \beta)}{( \tan\alpha + \sec\alpha)} \left[ 1 - (\frac{\pi}{4}-\frac{\alpha}{2})(\tan\alpha + \sec\alpha) \right] } \right)^2 }{ 8 \cos\alpha (\sin\alpha + \beta) 
      \left[ 1 - (\frac{\pi}{4}-\frac{\alpha}{2})(\tan\alpha + \sec\alpha) \right] }
  \label{eq:sstar_case1}
\end{eqnarray}
The above expression has to satisfy the constraint 
\begin{equation}
  0 < s^* < s_{c1}.
\end{equation}

For a perfect rectangular channel ($\alpha=0$), we obtain
\begin{equation}
  s^* = \frac{1}{\beta(8-2\pi)} \Big( 2+\beta - \sqrt{(2+\beta)^2 - \beta(8-2\pi)} \Big)^2.
\end{equation}
In this case, one can show $s^*<s_{c1}=(\frac{1}{2} - \frac{\pi}{8}) \beta$ is always satisfied.
Thus, for the rectangular cross-section, $s^*$ is always present in case (I) with four fingers. 

For a square cross-section ($\beta=2$), we obtain $s^*\simeq 0.06$. 
This is consistent with Refs~\cite{Dong1995, 2018_square}.

\subsection{Case (II)}

\begin{figure}[ht]
  \includegraphics[width=0.55\columnwidth]{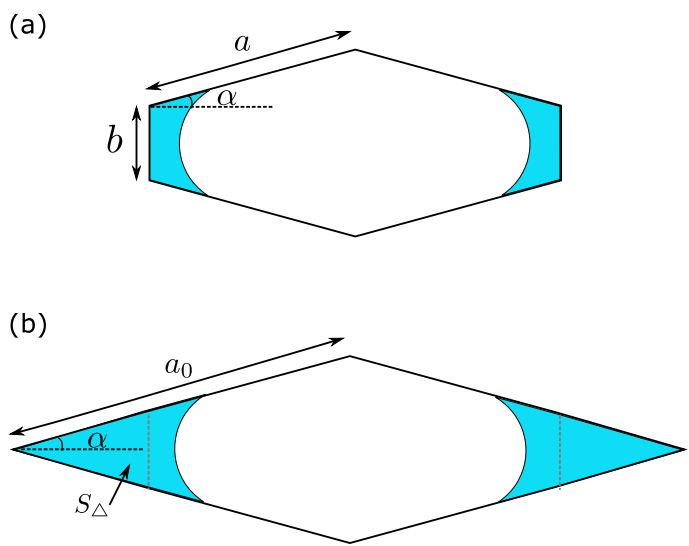}
  \caption{Demonstration of procedure to calculate the $s^*$ in case (II).
The meniscus position of $s^*$ is shown in (a). 
For a diamond channel $a_0 = a + b/(2\sin\alpha)$ (b) with the saturation $s_{\diamond}^*$, the position of the meniscus does not change in comparison with (a). }
  \label{fig:case2star}
\end{figure}

For case (II), the free energy density presented in Eq.~(\ref{eq:f_case2}) is written as a function of the intermediate variable $\ell$. 
The direct calculation of $f'(s)$ is cumbersome and leads to a complicated expression of $s^*$. 
Here we present a simpler derivation. 
Consider a diamond channel that is similar to the case (II) we are interested in.
In Fig.~\ref{fig:case2star}(a), the position of the meniscus corresponds to the $s^*$. 
It is easy to show that the same position of the meniscus in a diamond channel with $a_0 = a + b/(2\sin\alpha)$, shown in Fig.~\ref{fig:case2star}(b), will also correspond to the $s_{\diamond}^*$ for the diamond channel. 

In Appendix \ref{sec:diamond}, we derive an analytic expression of $s_{\diamond}^*$ for the diamond channel,
\begin{equation}
  \label{eq:sstar_diamond}
  s_{\diamond}^* = -1 + \frac{ 2 - 2 \sqrt{ 1 - \cos^2 \alpha \Big( 1 - (\frac{\pi}{2} - \alpha)\tan\alpha \Big) } }
                  { \cos^2 \alpha \Big( 1 - (\frac{\pi}{2} - \alpha)\tan\alpha \Big) }, 
\end{equation}
we can compute the $s^*$ for case (II) by the simple geometric calculation
\begin{equation}
  {S_{\diamond 0}} s^*_{\diamond} - 2 S_{\triangle} = S_0 s^*,
\end{equation}
where $S_0$ and $S_{\diamond 0}$ are the total area of the nearly-rectangular cross-section [Fig.~\ref{fig:case2star}(a)] and the diamond cross-section [Fig.~\ref{fig:case2star}(b)], respectively. 
$S_{\triangle}$ is the area of the truncated triangle. 
This leads to 
\begin{eqnarray}
  s^* &=& s^*_{\diamond} \frac{ (\sin\alpha + \frac{1}{2} \beta)^2}{ \sin\alpha ( \sin\alpha + \beta)} 
    - \frac{\beta^2}{ 4 \sin\alpha ( \sin\alpha + \beta )} \\
  &=& \frac{ ( \sin\alpha + \frac{1}{2}\beta )^2 
      \left[ -1 + \frac{ 2( 1 - \sqrt{ 1 - \cos^2\alpha [ 1 - (\frac{\pi}{2} - \alpha) \tan\alpha ] } )}
      { \cos^2 \alpha [ 1 - (\frac{\pi}{2} - \alpha) \tan\alpha ] } \right] - \frac{1}{4} \beta^2 }
      { \sin\alpha ( \sin\alpha + \beta )} .
      \label{eq:sstar_case2}
\end{eqnarray}
Similar to the case (I), the above expression is constrained by
\begin{equation}
  s_{c1} \le s^* \le s_{c2}.
\end{equation}

When $\alpha$ is close to zero (not equal to zero), the trigonometric functions take simple form $\sin\alpha \simeq  \alpha$, $\tan\alpha \simeq  \alpha$, $\cos\alpha \simeq 1$.
Using these simplifications, approximated expressions for the diamond and nearly-rectangle can be obtained 
\begin{equation}
  \label{eq:sstar_diamond_app}
  (s_{\diamond}^*)_{\rm app} =1-\sqrt{2\pi\alpha} , 
\end{equation}
\begin{equation}
  \label{eq:sstar_app}
  (s^*)_{\rm app} =1-\frac{(\beta+2\alpha)^2}{4\alpha(\alpha+\beta)}\sqrt{2\pi\alpha} .
\end{equation}
For the nearly-rectangular cross-section (\ref{eq:sstar_app}), the function is a non-monotonic function of $\alpha$. 
We can compute $\partial (s^*)_{\rm app} / \partial \alpha = 0$, leading to the extreme point
\begin{equation}
  (\alpha_{\rm max})_{\rm app} = \beta/3, \quad
  (s^*)_{\rm max,app} = 1 - \frac{49}{48} \sqrt{ 6 \pi \beta }. 
\end{equation}

\subsection{$s^*$ results}

We summarize the results of previous two sections. 
For a fixed value of $\beta$, the coexistence $s^*$ can locate either in the range of case (I) with menisci composed of four fingers, or case (II) with two fingers. 
For the four-finger case (I), the value of $s^*$ is given by Eq.~(\ref{eq:sstar_case1}) and the range of $s^*$ is $0 < s^* < s_{c1}$. 
For the two-finger case (II), the value of $s^*$ is given by Eq.~(\ref{eq:sstar_case2}) and the range is $s_{c1} \le s^* \le s_{c2}$.

\begin{figure}[ht]
  \includegraphics[width=0.9\columnwidth]{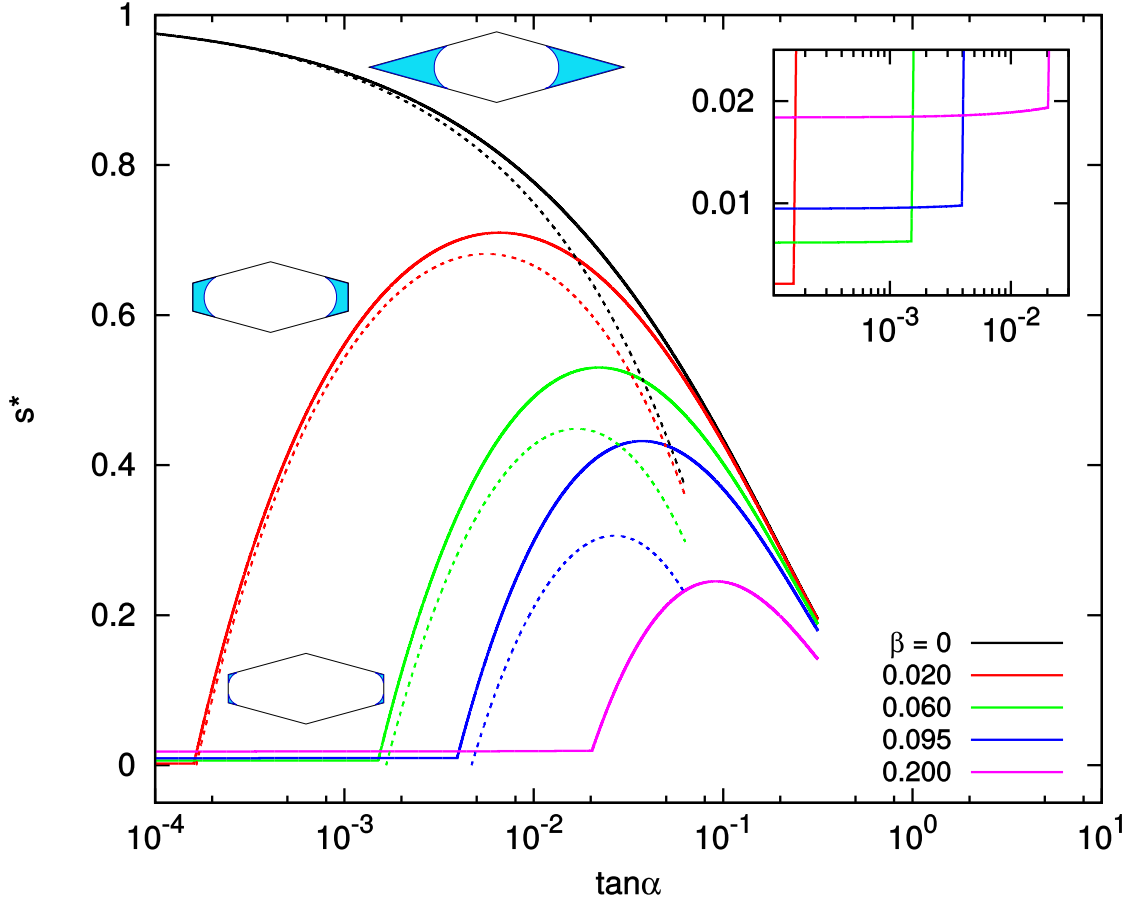}
  \caption{The $s^*$ value as a function of $\tan\alpha$ and $\beta$. 
The solid lines are exact solutions: Black, $\beta=0$, diamond channel, Eq.~(\ref{eq:sstar_diamond}); Red, $\beta=0.02$, nearly-rectangular channel, Eqs.(\ref{eq:sstar_case1}) and (\ref{eq:sstar_case2}); Green, $\beta=0.06$; Blue, $\beta=0.095$; Violet, $\beta=0.20$.
The dashed lines are approximated solutions given in Eqs.~(\ref{eq:sstar_diamond_app}) and (\ref{eq:sstar_app}). 
There is no approximated solution for $\beta=0.20$.
The inset shows the enlarged part near small $\tan\alpha$, where the transition from case (I) to case (II) takes place at $\tan \alpha_{\rm crit}$.}
  \label{fig:sstar}
\end{figure}

Figure \ref{fig:sstar} shows the value of $s^*$ as a function of $\tan\alpha$ for different value of $\beta$.  
For the special case of diamond channel, $\beta=0$, $s^*$ is given by Eq.~(\ref{eq:sstar_diamond}). 
This function is plotted with the solid black line in Fig.~\ref{fig:sstar}.
It is a monotonically decreasing function when $\tan\alpha$ increases.  
Also plotted is the approximated solution Eq.~(\ref{eq:sstar_diamond_app}). 
The approximated solution has good agreement with the exact formulas at small $\tan\alpha$, but shows noticeable deviation at large $\tan\alpha$.

For nearly-rectangular channel, $\beta > 0$, representative curves are shown in solid lines in Fig.~\ref{fig:sstar}.
The value of $s^*$ shows two different regions.  
When $\tan\alpha$ is small, $s^*$ corresponds to the four-finger configuration of case (I), and the expression is given in Eq.~(\ref{eq:sstar_case1}). 
In this case, the value of $s^*$ is also very small (see the inset for the absolute values).
When $\tan\alpha$ is large, $s^*$ is located in the range of case (II) and its expression is given by Eq.~(\ref{eq:sstar_case2}). 
In this case, the value of $s^*$ is large and has a non-monotonic dependence on $\tan\alpha$.
We also plot the approximated solution (\ref{eq:sstar_app}) in dashed lines. 
The agreement between the approximated and exact solutions is good when $\alpha$ is small or $\beta$ is small.  

\begin{figure}[ht]
  \includegraphics[width=1.0\columnwidth]{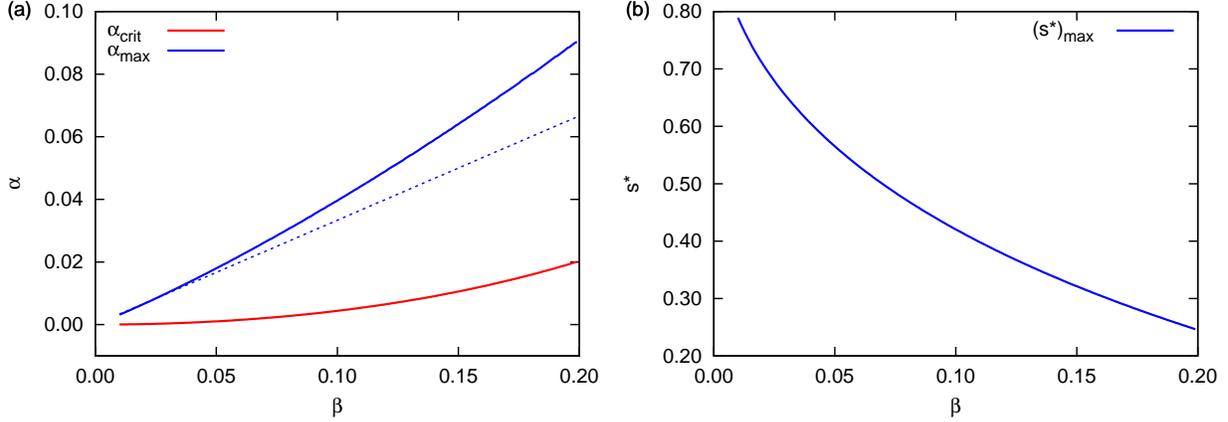}
  \caption{(a) The critical value $\alpha_{\rm crit}$ (red), when the location of $s^*$ changes from case (I) to case (II), is plotted as a function of the aspect ratio $\beta$. 
Also shown is the value of $\alpha_{\rm max}$ at which $s^*$ reaches the maximum. The blue solid line is the exact solution and the dashed blue line is the approximated solution $(\alpha_{\rm max})_{\rm app} = \beta/3$. (b) The maximum value of $(s^*)_{\rm max}$ is plotted as a function of $\beta$. }
  \label{fig:crit}
\end{figure}

The two regions of $s^*$ are connected at a critical $\alpha_{\rm crit}$ value. 
When $\beta$ increases, i.e. as the constriction becomes less elongated, the critical $\alpha$ moves to a larger value. 
This trend is also demonstrated in Fig.~\ref{fig:crit}(a) with the red solid line.

When $s^*$ is located in the range of case (II), the value of $s^*$ is very sensitive to $\alpha$. 
For example, for $\beta=0.02$ shown as the red line in Fig.~\ref{fig:sstar}, $s^*$ increases from nearly zero to about 0.7 when $\tan\alpha$ changes from $2 \times 10^{-4}$ to $5 \times 10^{-3}$ (less than 0.2$^{\circ}$ change).
In this range, the value $s^*$ increases first with increasing $\tan\alpha$, reaching a maximum, then decreases when $\tan\alpha$ increases further.
The location of the maximum $\alpha_{\rm max}$ increases with increasing $\beta$.
This is shown as the blue solid line in Fig.~\ref{fig:crit}(a). 
Also shown is the approximated value $(\alpha_{\rm max})_{\rm app} = \beta/3$ in dashed line, which again has better agreement with the exact solution at small $\beta$.
The maximum value of $(s^*)_{\rm max}$ is a decreasing function of $\beta$, shown in Fig.~\ref{fig:crit}(b).
When $\tan\alpha$ increases further after passing the maximum, the cross-section of the channel resembles a diamond with two tips removed (small $S_{\triangle}$ in Fig.~\ref{fig:case2star}). 
In this case, $s^*$ curves of different $\beta$ value all approach the curve of the diamond channel. 

In the experiment of Keita \emph{et al.} \cite{Keita2016}, the aspect ratio is {$\beta = 0.095$} and the deviation from the rectangle $\tan\alpha = 0.01$. 
From Fig.~\ref{fig:sstar}, we can obtain $s^*\simeq 0.2987$ by Eq.~(\ref{eq:sstar_case2}).  
This value of $s^*$ locates in case (II) and the menisci are composed of two fingers. 
Our result is consistent with the observation in Ref.~\cite{Keita2016} (see the left column of Fig.~\ref{fig:exp}).
The experimental value of $s^*$ \change{(at the finger/bulk interface)} is about 0.38, which is close to our prediction.

In the experiment of Seck \emph{et al.} \cite{Seck2018}, the aspect ratio is $\beta = 0.2$ and the cross-section is nearly perfect rectangle, so $\alpha$ is very small. 
In this case, Fig.~\ref{fig:sstar} shows that a very small $s^*$ exists over a large range of $\tan\alpha$ value, corresponding to the case (I) with four fingers. 
The value of $s^*$ given by Eq.~(\ref{eq:sstar_case1}) is $s^* \simeq 0.018$, again this is consistent with the observation in Ref.~\cite{Seck2018} (also see the right column of Fig.~\ref{fig:exp}).

\section{Conclusions}

In this paper, we studied the wetting phenomenon in a nearly-rectangular channel.  
In the equilibrium, the channel can be filled by a wetting fluid completely, corresponding to $s=1$, or partially filled with $s^* < 1$. 
The partial-wetting region is composed of either four or two fluid columns (referred as ``finger''). 
We analyzed different scenarios of wetting behaviors and obtained analytical results of $s^*$ as a function of the aspect ratio $\beta$ and the angle $\alpha$. 
We found
\begin{enumerate}
\item[(1)] When the cross-section is a nearly perfect rectangle, angle $\alpha$ is close to zero, the saturation $s^*$ located in the four-finger region and with a very small value. 

\item[(2)] When the angle $\alpha$ is large but still less than $1^{\circ}$ ($\tan\,1^{\circ} = 0.017$), the value of $s^*$ is very sensitive with respect to even a tiny change of $\alpha$. 
The meniscus takes the two-finger configuration. 
The value of $s^*$ can be as large as 0.8, which is significantly larger than that in the four-finger case. 
\end{enumerate}

\change{In this work, we have focused on the equilibrium case and discussed about the coexistence between the bulk region and finger region.
When the system has a slow dynamics or the Capillary number is small, our results of the saturation $s^*$ should apply at the interface between the bulk and the finger. 
We indeed found good agreement with the experimental observations \cite{Keita2016, Seck2018}. 
For simplicity, we have only discussed the case that the equilibrium contact angle is zero. 
The extension to a finite contact angle is straightforward. 
For fully wetting, a precursor film of molecular thickness might exist on the surface. 
In this case the free energy functions take the same forms, thus our prediction of $s^*$ should remain valid.}
An important and related problem is how the change of $s^*$ influences the imbibition dynamics \cite{2018_square, 2019_Taylor_rising, 2020_onethird}.
We expect {our framework} can provide theoretical support to the understanding of the wetting and drying in porous media, and in the design of micro/nanofluidic devices.

\begin{acknowledgments}
This work was supported by the National Natural Science Foundation of China (NSFC) through the Grant No. 21774004 (J.Z.), No. 91634202 and No. 11972212 (Q.S). 
M.D. acknowledges the financial support of the Chinese Central Government in the Thousand Talents Program. 
\end{acknowledgments}

\appendix
\section{Alternative derivation of $s^*$}
\label{sec:Mason}

In this appendix, we show the coexistence condition (\ref{eq:tangential}) is equivalent to the geometric condition suggested in Ref.~\cite{Mason1984}.

Mason and Morrow used the virtual work balance to derive a condition for the capillary pressure (Eq.~(4) in Ref.~\cite{Mason1984}).
Written in our notation, the condition is
\begin{equation}
  \label{eq:Mason}
  p_c (S_0 - S) = \gamma (P_{\rm S} \cos\theta_E + P_{\rm L}).
\end{equation}
Here $p_c$ is the capillary pressure. 
$S_0$ is the total area of the cross-section, $S$ is the area occupied by the fluid. 
$P_{\rm S}$ is the solid-vapor perimeter and $P_{\rm L}$ is the liquid-vapor perimeter. 

Using the definition of the capillary pressure (\ref{eq:pc}), we write the LHS of Eq.~(\ref{eq:Mason}) 
\begin{equation}
  \label{eq:Mason_LHS}
  p_c (S_0 - S) = - f'(s) \frac{S_0 - S}{S_0} = - f'(s) (1 - s).
\end{equation}
The free energy densities for the partial saturation and full saturation are
\begin{eqnarray}
  f(s) &=& (P_0 - P_{\rm S}) \gamma_{SL} + P_{\rm S} \gamma_{SV} + P_{\rm L} \gamma, \\
  f(1) &=& P_0 \gamma_{SL} ,
\end{eqnarray}
where $P_0$ is the length of the perimeter in the cross-section. 
$\gamma_{SL}$, $\gamma_{SV}$, and $\gamma$ are the interfacial tensions of solid-liquid, solid-vapor, and liquid-vapor interfaces, respectively. 
From the free energy densities, we calculate
\begin{equation}
  \label{eq:Mason_RHS}
  f(s) - f(1) = P_{\rm S} (\gamma_{SV} - \gamma_{SL}) + P_{\rm L} \gamma 
              = P_{\rm S} \gamma \cos\theta_E + P_{\rm L} \gamma .
\end{equation}
This is just the RHS of Eq.~(\ref{eq:Mason}). 

Combining Eqs.~(\ref{eq:Mason_LHS}) and (\ref{eq:Mason_RHS}), we obtain the same condition (\ref{eq:tangential}) for $s^*$
\begin{equation}
  f'(s) \Big|_{s=s^*} = \frac{ f(1) - f(s^*) }{ 1 - s^* } .
\end{equation}

\section{Diamond channel}
\label{sec:diamond}

\begin{figure}[ht]
  \includegraphics[width=0.5\columnwidth]{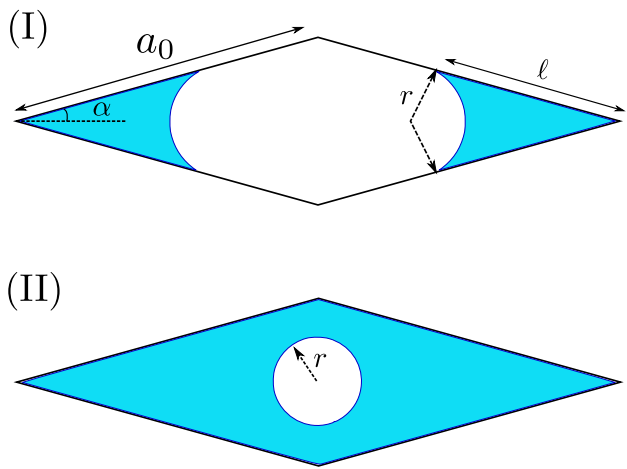}
  \caption{Two scenarios of the wetting fluid in the cross-section of a diamond channel. The side length of the diamond is $a_0$ and two adjacent sides form an angle $2\alpha$. (I) The liquid-vapor interfaces form two separated arcs in the corners. The radius of curvature of the arc is $r$ and the wetting length on the side is $\ell$. (II) The liquid-vapor interface is a full circle of the radius $r$.}
  \label{fig:diamond}
\end{figure}

In this appendix, we calculate $s^*$ for a channel with a diamond cross-section. 
The diamond has a side length $a_0$, and the sharp angle formed by the two sides is $2\alpha$. 
To calculate the free energy density $f(s)$, we need to distinguish the two cases shown in Fig.~\ref{fig:diamond}.

{\bf Case (I)}

Here the meniscus forms two separate fingers with a radius of curvature $r$.
The fluid covers a length $\ell$ on the side.
These two lengths satisfy the following relation
\begin{equation}
  r = \ell \tan \alpha . 
\end{equation}
The area of the fluid is given by
\begin{eqnarray}
  S &=& 2 \left( \ell^2 \tan \alpha - (\pi - 2 \alpha) \frac{1}{2} r^2 \right) \nonumber \\
    &=& 2 \ell^2 \tan \alpha \left( 1 - (\frac{\pi}{2} - \alpha) \tan \alpha \right).
\end{eqnarray}
The total area of the cross-section is
\begin{equation}
  {S_{\diamond 0}} = 2 a_0^2 \sin\alpha \cos\alpha.
\end{equation}
The saturation is then given by
\begin{equation}
  s = \frac{S}{{S_{\diamond 0}}} = \frac{1}{\cos^2 \alpha} \left( 1 - (\frac{\pi}{2} - \alpha) \tan \alpha \right) 
      \Big( \frac{\ell}{a_0} \Big)^2
\end{equation}
Since $\ell \le a_0$, the range of $s$ in case (I) is 
\begin{equation}
  0 < s \le s_{c2}, \quad s_{c2} = \frac{1}{\cos^2 \alpha} \left( 1 - (\frac{\pi}{2} - \alpha) \tan \alpha \right).
\end{equation}

The free energy density is 
\begin{equation}
  f(s) = \left( -4 + (2\pi - 4 \alpha) \tan \alpha \right) \ell \gamma
\end{equation}
\begin{equation}
  \label{eq:f_diamond}
  \frac{f(s)}{a_0 \gamma} = - \left( 4 \cos\alpha \sqrt{ 1 - (\frac{\pi}{2}-\alpha) \tan \alpha} \right)
  \sqrt{s} = -F_1 \sqrt{s},
\end{equation}
with
\begin{equation}
  F_1 = 4 \cos\alpha \sqrt{ 1 - (\frac{\pi}{2}-\alpha) \tan \alpha} .
\end{equation}

{\bf Case (II)}

In this case, the meniscus forms a complete circle with the radius $r$. 
The area of the fluid is
\begin{equation}
  S = {S_{\diamond 0}} - \pi r^2 .
\end{equation}
The saturation is 
\begin{equation}
  s = \frac{S}{{S_{\diamond 0}}} = 1 - \frac{\pi}{ \sin(2\alpha) } \Big(\frac{r}{a_0}\Big)^2 .
\end{equation}
Since $r \le a_0 \sin\alpha \cos\alpha$ (the shortest distance from the center to the side), the range of $s$ in case (II) is
\begin{equation}
  s_{c3} \le s \le 1, \quad s_{c3} = 1 - \frac{\pi}{4} \sin(2\alpha) .
\end{equation}

The free energy density is
\begin{equation}
  f(s) = - 4 a_0 \gamma + 2 \pi r \gamma,
\end{equation}
\begin{equation}
  \frac{f(s)}{a_0 \gamma} = - 4 + 2\pi \sqrt{ \frac{\sin(2\alpha)}{\pi} (1-s) }.
\end{equation}

For the complete saturation, the free energy density is 
\begin{equation}
  \label{eq:f1_diamond}
  \frac{f(1)}{a_0 \gamma} = -4 .
\end{equation}

{\bf Calculation of $s^*$} 

Now we can compute $s^*$.
Similar to the case of rectangular cross-section, $s_{\diamond}^*$ exists only in case (I).
We again use the condition (\ref{eq:tangential}) combined with the free energy functions (\ref{eq:f_diamond}) and (\ref{eq:f1_diamond})
\begin{eqnarray}
  \frac{\ud f}{\ud s} &=& - F_1 \frac{1}{2} \frac{1}{\sqrt{s}},  \\
  \frac{f(1)-f(s)}{1-s} &=& \frac{ -4 + F_1 \sqrt{s} }{1-s}.
\end{eqnarray}
The $s_{\diamond}^*$ is determined by the equation
\begin{equation}
  F_1 s - 8 \sqrt{s} + F_1 = 0 .
\end{equation}
The solution is
\begin{equation}
  s_{\diamond}^* = \left( \frac{ 4 - \sqrt{16-F_1^2} }{F_1} \right)^2 
      = -1 + \frac{ 2 - 2 \sqrt{ 1 - \cos^2 \alpha \Big( 1 - (\frac{\pi}{2} - \alpha)\tan\alpha \Big) } }
                  { \cos^2 \alpha \Big( 1 - (\frac{\pi}{2} - \alpha)\tan\alpha \Big) }.
\end{equation}

\section{Equivalence of coexistent conditions in a diamond channel and a nearly-rectangular channel}
\label{sec:eq_coex}

Here we show the equivalence of the coexistent conditions in a nearly-rectangular channel [Fig.~\ref{fig:case2star}(a)] and a diamond channel [Fig.~\ref{fig:case2star}(b)].
We will denote quantities in the diamond channel with a subscript $\diamond$.

The free energy functions in these two channels are related by 
\begin{equation}
  \label{eq:fs_d}
  f(s) = f_{\diamond}(s_{\diamond}) + \frac{2b}{\sin\alpha} \gamma - 2b \gamma .
\end{equation}
The area occupied by the fluid in two channels are given by
\begin{equation}
  \label{eq:S_d}
  S = S_{\diamond} - 2 S_{\triangle}.
\end{equation}
These two equations lead to the following relation
\begin{equation}
  \label{eq:dfds_d}
  \frac{ \ud f(s) }{\ud S} = \frac{ \ud f_{\diamond} }{ \ud S_{\diamond} }.
\end{equation}
Using the definition of the saturation $s=S/S_0$ and $s_{\diamond} = S_{\diamond}/S_{\diamond 0}$, we can rewrite Eq.~(\ref{eq:dfds_d}) into 
\begin{equation}
  \label{eq:coex_d1}
  \frac{ \ud f(s) }{\ud s} = \frac{ \ud f_{\diamond} }{ \ud s_{\diamond} } \frac{S_0}{S_{\diamond 0}}.
\end{equation}

From Eqs.~(\ref{eq:fs_d}) and (\ref{eq:S_d}), we see that the free energy and the area for two cases differ only by a constant term, thus
\begin{eqnarray}
  f(1) - f(s) &=& f_{\diamond}(1) - f_{\diamond}(s_{\diamond}) , \\
  S_0 (1-s) &=& S_{\diamond 0} ( 1-s_{\diamond} ).
\end{eqnarray}
These lead to 
\begin{equation}
  \label{eq:coex_d2}
  \frac{ f(1) - f(s) }{ 1-s } = \frac{ f_{\diamond}(1) - f_{\diamond}(s_{\diamond})}{ 1 - s_{\diamond}} \frac{S_0}{S_{\diamond 0}}.
\end{equation}

When the saturation in the diamond channel satisfies the coexistent condition (\ref{eq:tangential}),
\begin{equation}
  \frac{\ud f_{\diamond} (s_{\diamond}) }{\ud s_{\diamond}} \Big|_{s_{\diamond} = s^*_{\diamond}} 
  = \frac{ f_{\diamond}(1) - f_{\diamond}(s^*_{\diamond}) }{ 1-s^*_{\diamond} },
\end{equation}
we immediately obtain from Eqs. (\ref{eq:coex_d1}) and (\ref{eq:coex_d2}) that the corresponding saturation in the nearly-rectangular channel 
also satisfies the coexistent condition
\begin{equation}
  \frac{ \ud f(s) }{ \ud s} \Big|_{s=s^*} = \frac{ f(1) - f(s^*)}{1-s^*}.
\end{equation}

\bibliography{wetting}

\end{document}